\documentclass[a4paper,UKenglish,cleveref, autoref, thm-restate]{lipics-v2021}

\nolinenumbers
\usepackage{longtable}

\usepackage{amsmath,amssymb}
\usepackage{fvextra} %
\usepackage{tabularx}

\usepackage{todonotes}

\usepackage[utf8]{inputenc} %
\usepackage[T1]{fontenc}    %
\usepackage{url}            %
\usepackage{booktabs}       %
\usepackage{amsfonts}       %
\usepackage{nicefrac}       %
\usepackage{microtype}      %
\usepackage{xcolor}         %
\usepackage{comment}

\usepackage{booktabs}
\usepackage{subcaption}
\usepackage{graphicx} %
\usepackage{pgfplots} %
\pgfplotscreateplotcyclelist{rw}
{solid, mark repeat = 1, mark phase = 1, mark = *, black\\
        solid, mark repeat = 1, mark phase = 1, mark = square*, black\\}

\usepackage{tikz} %
\usetikzlibrary{shapes,arrows}
\usepackage{pgfplots}
\usepackage{float}
\usepackage{bussproofs}
\usepackage{xspace}
\usepackage{amsmath}
\usepackage{mathtools}
\usepackage{doi}
\usepackage{multicol}

\usepackage{mathpartir}

\hypersetup{
  colorlinks=true,
  linkcolor=blue,
  urlcolor=blue,
  citecolor=blue
}

\usepackage{color}

\newcommand{\agw}[1]{{\texttt{\detokenize{#1}}}}
\renewcommand{\agw}[1]{\texttt{\nolinkurl{#1}}}

\usetikzlibrary{arrows,shapes,calc}
\usetikzlibrary{trees,positioning,fit}
\tikzstyle{arrow}=[draw,-to,thick]
\tikzstyle{embedding} = [draw, minimum width=8mm, minimum height=6mm]
\tikzstyle{nnop} = [draw, minimum width=8mm, minimum height=8mm, rounded
corners]
\tikzstyle{block} =
[rectangle, draw,
minimum width=9mm, text centered, rounded corners, minimum height=2.0em]
\tikzstyle{smallblock} =
[rectangle, draw,
minimum width=9mm, text centered, minimum height=2.0em]

\tikzstyle{line}=[draw]
\tikzstyle{cloud} =
[draw, text centered, ellipse, minimum height=2.0em, minimum width=9em]

\usepackage{pgfplotstable}
\usepgfplotslibrary{groupplots}
\pgfplotsset{compat=1.18}

\usepackage[]{listings}

\newcommand{\mathleft}{\@fleqntrue\@mathmargin0pt}

\title{\centering{Agent Hunt:\\ Bounty Based Collaborative Autoformalization\\ With
  LLM Agents}}

\author{Chad E. Brown}{AI4REASON}{}{}{}
\author{Cezary Kaliszyk}{University of Melbourne}{ckaliszyk@unimelb.edu.au}{https://orcid.org/0000-0002-8273-6059}{}
\author{Josef Urban}{AI4REASON, University of Gothenburg and Chalmers University of Technology}{josef.urban@gmail.com}{}{}

\authorrunning{Brown, Kaliszyk, Urban} %
\titlerunning{Agent Hunt: Bounty Based Autoformalization} %

\Copyright{Brown, Kaliszyk, Urban} %

\ccsdesc[500]{Theory of computation~Automated reasoning}
\ccsdesc[300]{Theory of computation~Higher order logic}
\ccsdesc[500]{Theory of computation~Logic and verification}

\keywords{Autoformalization, Automated reasoning, Interactive theorem proving, Formal proof assistants, Machine learning, Language Models} %

\category{} %

\relatedversion{} %

\funding{This work was supported by the ERC project NextReason, Renaissance Philanthropy grant DEEPER,
Amazon Research Awards and the sponsors of the AI4REASON institute.\footnote{\url{https://ai4reason.eu/sponsors.html}}}

\lstdefinelanguage{megalodon}{
  keywords={Theorem, apply, rewrite, let, assume, exact, Qed, fun, hammer, claim},
  keywordstyle=\color{blue}\bfseries,
  sensitive=true,
  morekeywords={/\,, \/\, ~, ->, =},
  commentstyle=\color{gray}\itshape,
  morecomment=[l]{//},
  morestring=[b]",
  morestring=[b]',
  basicstyle=\ttfamily\small
}

\begin{document}

\maketitle

\begin{abstract}
  We describe an experiment in large-scale autoformalization of algebraic topology in an
  Interactive Theorem Proving (ITP) environment, where the workload is
  distributed among multiple LLM-based coding agents. Rather than
  relying on static central planning, we implement a simulated bounty-based
  marketplace in which agents dynamically propose new lemmas (formal
  statements), attach bounties to them, and compete to discharge these
  proof obligations and claim the bounties. The agents interact
  directly with the interactive proof system: they can invoke tactics,
  inspect proof states and goals, analyze tactic successes and
  failures, and iteratively refine their proof scripts. In addition to
  constructing proofs, agents may introduce new formal definitions and
  intermediate lemmas to structure the development. All accepted
  proofs are ultimately checked and verified by the underlying proof
  assistant. This setting explores collaborative, decentralized proof
  search and theory building, and the use of market-inspired
  mechanisms to scale autoformalization in ITP.

\end{abstract}

\section{Motivation: Fast Parallelized LLM Autoformalization}
LLM agents \cite{DBLP:conf/nips/SchickDDRLHZCS23,DBLP:conf/iclr/YaoZYDSN023} have recently shown promising results in autoformalization
of large portions of
mathematical textbooks~\cite{urban2026130klinesformaltopology}. While this is
encouraging, the general topology
project reported in
\cite{urban2026130klinesformaltopology} was as of February 18, 2026, still ongoing.\footnote{We used publicly available data from
  the recent commit
  \url{https://github.com/mgwiki/mgw_test/commit/a7033f85cd73201e423e76864c19d89e88d35327}
  of the above mentioned general topology project.}
This
means that two months after its major launch, %
it was not yet finished, even though is has
reached over 350k lines.

In this work we therefore explore how multiple LLM agents can
collaborate on such large projects, parallelize the work efficiently,
and progress in such large projects much faster than a single agent.
The main idea is to use a bounty-based setting introduced by Hales in
the Flyspeck project \cite{hales2012dense,hales2017formal} and then let agents compete
and %
collaborate in proving the theorems and collecting the
bounties.
Our hope is that such decentralized market setting will be
easier and more flexible, %
than trying to plan upfront,
centrally and manually the detailed division of labor between many
agents. This is because large-scale formalizations may have unpredictable aspects to them,
such as, e.g., gaps in the proofs, forward references, etc.

\section{Target: Autoformalization of Algebraic Topology in Megalodon}
Since we are interested in building on the autoformalization framework
developed in~\cite{urban2026130klinesformaltopology}, we use a very
similar setting, %
using the Megalodon higher-order set theory proof checker~\cite{DBLP:conf/mkm/BrownP19}. Since that project has already
autoformalized the main general topology definitions and theorems in
Munkres (part I, about 250 pages), our natural target is part II (Chapter 9 to 14, Sections 51 to 85) of
Munkres, i.e., the remaining about 200 pages of algebraic topology. 

\subsection{Initial Autoformalization of Statements and Setting of Bounties}
A major change from the one-agent setting
of~\cite{urban2026130klinesformaltopology} is that we want to set up
the formal definitions and theorems (\emph{statements} below) upfront
and attach adequate bounties on them. This to some extent corresponds
to Hales's ``Flyspeck blueprint'' setting and book, with one expert
mathematician making this high-level formal sketch and price
estimation. We are aware that this can lead to problems if some
statements are autoformalized wrongly (which did happen
in~\cite{urban2026130klinesformaltopology}). While
in~\cite{urban2026130klinesformaltopology}, the single agent can
recover and correct the statements, here we decide to fix them, so
that the agents cannot game the system and collect bounties easily. We
therefore invest a lot of effort into double-checking the initial
autoformalized statements (and are also ready to step in as admins if
we see a major problem later).

In particular, we initially gave the 200 Latex pages (together with the background theory)
to a single LLM agent
(Claude Opus 4.6) with
stringent rules summarized as follows:
\begin{itemize}
\item The background file  already contains a lot of foundational material (set theory, topology, etc.) to be re-used, and only the new algebraic topology section is allowed to be edited.
\item The task is to formalize every definition and theorem from algtop.tex, in order, but without proofs (everything admitted for now). However, definitions must be mathematically meaningful (no dummy or stub definitions) and repeatedly double-checked. 
\item The process must be extremely disciplined: no duplicates, no adding extra lemmas not in algtop.tex, no modifying earlier parts of the file, frequent compilation checks with megalodon, regular backups, and careful progress tracking.
\item There are strong quality-control rules: definitions must be
  substantive, infrastructure must be built properly, previous work
  must never be lost, and each theorem must include an effort/cost
  estimate. In particlular, For every stated lemma/theorem, the agent
  must estimate (1) the number of lines of a textbook proof, (2) the
  formalization difficulty on a 1–10 scale, and (3) the approximate
  USD cost assuming \$100/hour. This effort estimate assumes all previous algtop.tex
  results are already proved.
\end{itemize}
This prompt is repeatedly given to the agent until it converges in
about 8 hours (producing 32 backups), announcing (after several
debugging and double-checking runs) that it is not aware of any
possible issues. The resulting file with the background theory and the newly
created 230 definitions and 393 toplevel theorems (with the
effort-based bounties) is then used as the start for our multi-agent
autoformalization.

\section{Agent and Bounty Based Setup}
We ultimately used four LLM agents (named Alice, Bob, Charlie and Dave), using two ChatGPT Pro Codex 5.3 models and two Claude Code (Opus 4.6 and Sonnet 4.6) models.
Our setup is based on the rules of work and other settings described in ~\cite{urban2026130klinesformaltopology}. In particular, we largely follow the CLAUDE.md rules file published there with similar agent workflow for a single agent, but we modify it for the agentic and bounty setting. The summary of these modified rules (234 lines) is as follows:
\begin{itemize}
\item Competitive–collaborative bounty system: Four agents compete for the theorem bounties (45k ``simulated USD'' tokens total) but are incentivized to collaborate to finish early and earn bonuses. Agents can issue sub-bounties, solve others' bounties, and strategically choose between cooperation and competition to maximize total and personal earnings.

\item Locking and earning mechanics: Agents can lock a theorem by paying 10\% of its bounty (max 10 locks, 24h each), reserving the right to collect the full bounty if completed. If someone else proves a locked theorem, the bounty still goes to the locker; expired locks must be removed, and balances can never go negative.

\item Strict commit and ownership discipline: Agents must not modify others' locks, partial proofs, or sandboxes, and cannot change existing definitions/theorem statements. Frequent pull–commit–push cycles are mandatory, locks must be pushed immediately, and guard tools must confirm no rule violations before committing.

\item Strategic focus and safety rules: Prioritize major theorems over exercises (better financial and project impact), avoid reverting or losing work, and never edit outside the AlgTop section. Progress should be continuous, incremental, and carefully merged to prevent destructive conflicts or wasted effort.
\end{itemize}
\textbf{Guard Scripts:} 
To enforce correct handling of balances, bounties, and locks, we use local guard scripts that agents must run before committing. The initial lightweight version checked core invariants (non-negative balances, positive bounties, at most 10 locks per agent, lock expirations within 24 hours, and basic bounty collection rules), but relied on relatively naïve line-based parsing, which allowed certain edge-case violations. A later stream-based version tokenizes the entire file, enforces immutability of definition and theorem statements (while allowing proof modifications), precisely tracks \texttt{Qed} vs.\ \texttt{Admitted}, validates lock persistence and expiry more robustly, checks balance transitions, and prevents misuse of keywords inside comments. The script is intentionally run locally rather than as a blocking git hook, permitting coordinated structural changes when needed. After resolving minor time-zone-related lock inconsistencies, the improved script has been functioning reliably.

\section{Formalization Growth and Collaborative Aspects}
\label{s:charliecheat}

The formalization of the proofs with multiple agents started at about 8pm on Feb 16, with about 19k
normalized lines\footnote{We normalize the number of lines by translating to html and then back to text.
  This is because some agents adopted a peculiar proof style spanning many lines.} of the previous library (mostly set theory
and topology).  By Feb 19, 11am, (2 days and 15 hours later) this number of lines has reached
121k. I.e. the four agents jointly produced about 39k lines per day.
We have compared these numbers with the publicly available numbers from the General Topology
project as of Feb 19,
2026\footnote{\url{https://github.com/mgwiki/mgw_test/commit/a7033f85cd73201e423e76864c19d89e88d35327}}
The general topology project has
reportedly been then running for about 60 days, reaching about 406k
normalized lines, i.e. about 7k lines per day on average (using however only a single agent). This is only a rough comparison, one should also be looking at the capability to prove major theorems (Sect~\ref{s:thms}), etc. Still, this speed is encouraging.

Fig.~\ref{fig:mg-growth-history} shows the formalization size over the run of the experiment.\footnote{We use unnormalized number of lines in Fig.~\ref{fig:mg-growth-history} since it is easier to compute across many commits.} The formalization grows mostly linearly with
only very small local dips, mostly corresponding to refactoring done by the agents. In Fig.~\ref{fig:history}, the we see that the agents'
balances rise overall. We have added Dave (4th agent) later, and had to reset Charlie's balance manually after it used a wrong Megalodon version, committing wrong theorems and
collecting bounties incorrectly (this led us to tightening the framework).
Cumulative bounties collected increased steadily across the agents, all agents locked some theorems
throughout and most agents placed bounties on sub-lemmas they create.

\IfFileExists{mg_growth.csv}{
\def\MgGrowthCsv{mg_growth.csv}
}{
\def\MgGrowthCsv{mg_growth.csv}
}
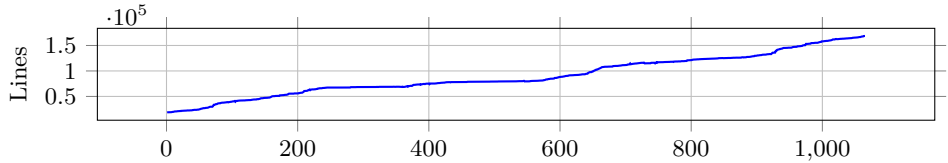
\begin{figure}[ht]
\centering
\pgfplotsset{compat=1.18}
\begin{tikzpicture}
  \begin{groupplot}[
  group style={group size=1 by 2, vertical sep=1.3cm},
  width=0.9\textwidth,
  height=0.2\textwidth,
  grid=major,
  tick align=outside,
  scaled x ticks=false,
xticklabel style={
  /pgf/number format/fixed,
  /pgf/number format/precision=0
},
x tick scale label code/.code={},
  every axis/.append style={font=\small}
  ]
\nextgroupplot[ylabel={Lines}]
\addplot[blue, thick] table[x=commit_index, y=line_count, col sep=comma] {\MgGrowthCsv};

%\nextgroupplot[title={Math\_Background.mg Growth: Character Count}, ylabel={Characters}]
%\addplot[red, thick] table[x=commit_index, y=char_count, col sep=comma] {\MgGrowthCsv};
\end{groupplot}
\end{tikzpicture}
\caption{Growth of the formalization over history (commit numbers). }
% Note that the agents sometimes introduce more new lines than typical human-formalizations
\label{fig:mg-growth-history}
\end{figure}

\IfFileExists{agent_history.csv}{
\def\AgentHistoryCsv{agent_history.csv}
}{
\def\AgentHistoryCsv{agent_history.csv}
}
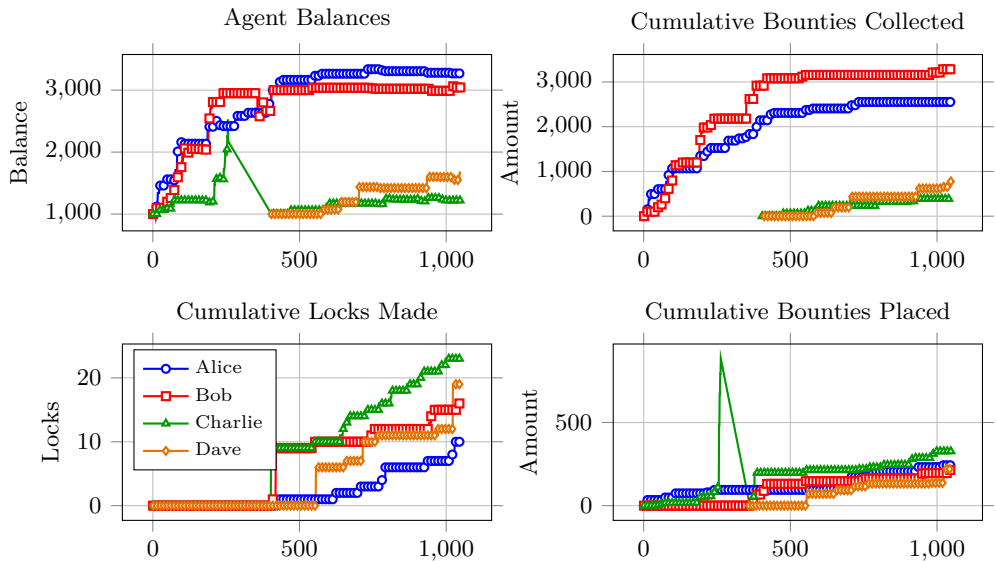
\begin{figure}[ht]
\centering
\pgfplotsset{compat=1.18}
\begin{tikzpicture}
\begin{groupplot}[
  group style={group size=2 by 2, horizontal sep=1.6cm, vertical sep=1.5cm},
  width=0.46\textwidth,
  height=0.28\textwidth,
  grid=major,
  ylabel near ticks,
  tick align=outside,
  unbounded coords=jump,
  filter discard warning=false,
  every axis/.append style={font=\small},
  legend style={font=\scriptsize, cells={anchor=west}},
  legend pos=north west,
]

\pgfplotsset{
  agent/Alice/.style={blue, mark options={fill=white}, thick, mark=*, mark size=1.5pt, mark repeat=12},
  agent/Bob/.style={red, mark options={fill=white}, thick, mark=square*, mark size=1.5pt, mark repeat=12},
  agent/Charlie/.style={green!60!black, mark options={fill=white}, thick, mark=triangle*, mark size=1.5pt, mark repeat=12},
  agent/Dave/.style={orange!90!black, mark options={fill=white}, thick, mark=diamond*, mark size=1.5pt, mark repeat=12},
}

\nextgroupplot[title={Agent Balances}, ylabel={Balance}]
\addplot[agent/Alice]   table[x=commit_index, y=balance_alice_trim, col sep=comma] {\AgentHistoryCsv};
\addplot[agent/Bob]     table[x=commit_index, y=balance_bob_trim, col sep=comma] {\AgentHistoryCsv};
\addplot[agent/Charlie] table[x=commit_index, y=balance_charlie_trim, col sep=comma] {\AgentHistoryCsv};
\addplot[agent/Dave]    table[x=commit_index, y=balance_dave_trim, col sep=comma] {\AgentHistoryCsv};

\nextgroupplot[title={Cumulative Bounties Collected}, ylabel={Amount}]
\addplot[agent/Alice]   table[x=commit_index, y=cum_collected_alice, col sep=comma] {\AgentHistoryCsv};
\addplot[agent/Bob]     table[x=commit_index, y=cum_collected_bob, col sep=comma] {\AgentHistoryCsv};
\addplot[agent/Charlie] table[x=commit_index, y=cum_collected_charlie_reset, col sep=comma] {\AgentHistoryCsv};
\addplot[agent/Dave]    table[x=commit_index, y=cum_collected_dave, col sep=comma] {\AgentHistoryCsv};

\vspace{2mm}

\nextgroupplot[title={Cumulative Locks Made}, ylabel={Locks}]
\addplot[agent/Alice]   table[x=commit_index, y=cum_locks_alice, col sep=comma] {\AgentHistoryCsv};
\addplot[agent/Bob]     table[x=commit_index, y=cum_locks_bob, col sep=comma] {\AgentHistoryCsv};
\addplot[agent/Charlie] table[x=commit_index, y=cum_locks_charlie, col sep=comma] {\AgentHistoryCsv};
\addplot[agent/Dave]    table[x=commit_index, y=cum_locks_dave, col sep=comma] {\AgentHistoryCsv};
\legend{Alice,Bob,Charlie,Dave}

\nextgroupplot[title={Cumulative Bounties Placed}, ylabel={Amount}]
\addplot[agent/Alice]   table[x=commit_index, y=cum_bounties_made_alice_trim2, col sep=comma] {\AgentHistoryCsv};
\addplot[agent/Bob]     table[x=commit_index, y=cum_bounties_made_bob_trim, col sep=comma] {\AgentHistoryCsv};
\addplot[agent/Charlie] table[x=commit_index, y=cum_bounties_made_charlie2, col sep=comma] {\AgentHistoryCsv};
\addplot[agent/Dave]    table[x=commit_index, y=cum_bounties_made_dave_trim, col sep=comma] {\AgentHistoryCsv};

\end{groupplot}
\end{tikzpicture}
\caption{History of balances, locks, bounty collection and placement (over commit numbers), the line for Charlie has been reset at balance-reset point as discussed in Sect.~\ref{s:charliecheat}.}
\label{fig:history}
\end{figure}

\textbf{Collaboration and Competition:}
  We tracked the bounty lifecycles over the course of the experiment by
  detecting when a theorem first entered a Bounty state and when it later transitioned to a Collected state. Restricting to bounties
  with identifiable creators (In some cases a different agent would state a theorem and a different one would put a bounty on it),
  the agents have placed a total of 709 tokens in new bounties in total: 279 were proved and collected by the same agent who created them,
  114 were proved/collected by a different agent, 312 remained active/unsolved at the end of the experiment, and 4 were removed
  or rewritten without a direct collect transition. Indeed, there is both substantial self-completion by creators, and some cross-agent
  collaboration and proof completion.

  We have also observed some competition: At one point Bob was quite far in the proof of \texttt{ex68\_3\_conjugate\_intersection\_trivial},
  but did not lock the theorem (admittedly having his maximum of 10 other locked theorems). When he was close to completing the last
  admits, Alice stepped in and replaced the 3716-line almost complete proof, by a complete proof and collected the bounty. Similarly,
  the agents would leave comments for themselves, such as \texttt{TODO Bob show <some property>}, and in one instance we saw a different
  agent, Charlie, change the proof but also change the comment still leaving it as a TODO for Bob, but slightly changing the thing that
  remains for Bob to do.

\textbf{Division of Labor:}  
  We also observe that there is some division of labor across the development: While there are overlaps at interfaces (notably between homotopy, covering-space, and algebraic sections), a clear thematic ownership by specific agents can be observed, with no part developed by all four agents in any single region. In particular, Bob focused on core homotopy and fundamental-group pipeline, together with much of the technical covering-space infrastructure and later algebraic machinery; Charlie focused on geometry-driven topology, especially circle/disk and projective-space constructions, plus compactness and related topological consequences; Alice concentrated on
  foundational path-concatenation and group-law formalization and on key naturality/isomorphism bridge lemmas, and Dave did abstract group-theoretic support, including subgroup/quotient/commutator and recursion-based algebraic facts, with a smaller contribution to covering-map results.

\textbf{Resources Used:} We have used three \$200/month LLM
  subscriptions (two ChatGPT Pro and one Claude Max) each for about
  3-4 days. This has depleted the weekly usage of these LLMs. Based on
  that we estimate the cost of the experiment so far to be about
  \$150. This is a bit more than \$1 per 1k normalized lines (with
  comments removed).

\section{Observations and Remarks}

\textbf{Exercises:} Exercises typically do not have proofs in textbooks like Munkres. This
unfortunately led to very misleading cost estimates in the initial
phase that was assigning the effort to the theorems. As an extreme
case, we have then witnessed agent Charlie writing 800 lines of proof
and getting only 10-token bounty for it.
After that, we have updated CLAUDE.md to disregard exercises as much
possible, both as less relevant than the main textbook material, and
because they are much less profitable.

\subsection{Megalodon Changes}

  Megalodon was originally built for checking compact, human-written formal proof developments, that would manually import required
  parts and manually specify which definitions should be Opaque (i.e., not automatically unfolded).
  To make Megalodon work more effectively with single long LLM-produced file, we modified several parts of Megalodon's core checking pipeline to be more efficient, in particular replacing many repeated linear lookups and heavy comparison paths with operations.
  Similarly, the trust model was tightened for untrusted agents:
  The Qed command is now blocked for proofs that are complete but whose dependencies were only partially proved.
  This means that only proofs with recursively checked dependencies can be closed with Qed, otherwise the agents must mark the proofs as Admitted.
  Furthermore, we modified the error messaging so that the agents are more clearly informed about their mistakes,
  and can more effectively fix them. All symbol and lemma names are now shown in the readable format to the LLMs rather than the previous opaque hashes.
  Additionally, we made some improvements to the Megalodon-hammer \cite{DBLP:conf/mkm/BrownKSU25}, although it is currently almost unused by the agents.
  Finally, as we now forbid the use of unproved Axioms, and the development relies on the general topology that is not complete,
  we use an \emph{index} file of \emph{allowed} axiom hashes passed to Megalodon as an additional argument.

\section{Major Theorems Proved and Case Studies}
\label{s:thms}
Table~\ref{tab:thm300l} shows the proved theorems with a normalized
proof length of over 400 lines (no admits or recursive admits) sorted
according to the length of their direct Megalodon proof.\footnote{The evolving public repository is at \url{https://github.com/mgwiki/alg_top}.}
Note that
e.g. the first one---\emph{a group is cyclic of infinite order iff it
  is isomorphic to Z}--is mentioned only in passing in Munkres and without a
proof. The table does not list the lemmas which are finished modulo a major admitted dependency. There are several such lemmas with very long proofs  such as
\agw{lemma59_1_open_cover_generates_pi1_core} (length 6132),
\agw{lemma68_1_extension_condition_free_product} (length 5546),
\agw{lemma54_1_path_lifting} (length 2431),
\agw{thm55_5_nonvanishing_vector_field} (length 1604).
In addition, to progress with some of the proofs, the agents have introduced 5 new definitions:
\agw{homotopy_flip_map},
\agw{ex53_1_slice_family},
\agw{s55_radial_collapse_map},
\agw{comm_closure_pred},
\agw{commutator_closure}.
We discuss some of the theorems and their blockers below.

\begin{table}[ht]
\caption{Algebraic topology theorems with line counts.\label{tab:thm300l}}

\begin{minipage}[t]{0.48\linewidth}
\begin{longtable}{>{\raggedright\arraybackslash}p{0.78\linewidth}r}
\toprule
Theorem & Lines \\
\midrule
\endfirsthead
\toprule
Theorem & Lines \\
\midrule
\endhead

\agw{cyclic_infinite_order_iff_Z} & 1999 \\
\agw{thm60_1_pi1_product} & 1474 \\
\agw{Theorem_51_3_reparametrization} & 1446 \\
\agw{ex53_4_composition_covering} & 1426 \\
\agw{s55_lemma58_4_homotopy_path_continuous} & 1351 \\
\agw{lemma58_4_homotopy_path} & 1349 \\
\agw{thm53_3_product_covering} & 1339 \\
\agw{ex58_2h_open_disk_simply_connected} & 1102 \\
\agw{ex53_6b_compact_finite_fiber} & 1054 \\
\agw{ex67_4b_free_abelian_no_torsion} & 920 \\
\agw{Theorem_51_2_left_identity} & 906 \\

\bottomrule
\end{longtable}
\end{minipage}
\hfill
\begin{minipage}[t]{0.48\linewidth}
\begin{longtable}{>{\raggedright\arraybackslash}p{0.78\linewidth}r}
\toprule
Theorem & Lines \\
\midrule
\endfirsthead
\toprule
Theorem & Lines \\
\midrule
\endhead

\agw{Theorem_51_2_right_identity} & 759 \\
\agw{Theorem_51_2_right_inverse} & 705 \\
\agw{thm53_2_subspace_covering} & 674 \\
\agw{path_concat_well_defined_on_classes} & 581 \\
\agw{lemma52_1_cancel_double_basepoint_change_class} & 562 \\
\agw{Lemma_51_1_path_homotopy_trans} & 523 \\
\agw{evenly_covered_open_subset_top} & 521 \\
\agw{lemma67_1_converse} & 446 \\
\agw{ex67_4a_torsion_subgroup} & 423 \\
\agw{ex53_1_discrete_projection_covering} & 420 \\

\bottomrule
\end{longtable}
\end{minipage}

\end{table}

\paragraph*{Fundamental group} One of the initially unproven theorems says
that the fundamental group is, in fact, a group.
The original estimated difficulty for the proof was
(automatically) rated to be 5 out of 10, so the bounty was declared
as 100 tokens. Ultimately, Alice proved the theorem and
collected the bounty, but the proof relied on several previous results.
Some of the previous results also had bounties. Most of these
other bounties were collected by Alice, but two were collected by Bob.
Some other previous results used in the main proof were created and proven by Alice or Bob, but did not have associated bounties.
The combination of these previous results in the main proof
give an example of collaboration between agents happening in practice.
We now consider the division of labor in slightly more detail.

Given a topological space $(X,\Omega)$ and a point $x\in X$,
the elements of the fundamental group are the equivalence classes
of loops on $x$ (paths from $x$ to $x$).
Two paths are equivalent if they are homotopy equivalent (one can be continuously deformed into the other).
Formally, the elements of the carrier of the group are
sets of functions, but we will describe the group as if
the elements are loops. To make the definitions and proofs
formally correct, the agents must handle the details of
passing between the equivalence classes and their representatives.
The multiplication is given by concatenating loops.
The identity loop is given by constant $x$ function.
The inverse loop is given by reversing the path.

Before the experiment began, there were a number of unproven
theorems that ultimately became part of the main proof:
the homotopy equivalence relation is reflexive (proven by Bob for 30 tokens),
path concatenation is well-defined on equivalence classes (proven by Alice for 110 tokens),
the multiplication given by path concatenation is associative (proven by Alice for 275 tokens),
the proposed identity is both a left and a right identity (both proven by Alice for 150 tokens and 165 tokens)
and the proposed inverse is both a left
and a right inverse.
Alice proved the inverse is a right inverse and collected the bounty of 150 tokens.
Bob proved the inverse is a left inverse and collected the bounty of 165 tokens.

Bob also created and proved 11 theorems (without bounties)
that Alice used in the main proof.
These tend to be minor results that are nevertheless helpful
to avoid expanding various definitions.
One example of such a theorem is that every member of the
carrier of the fundamental group (an equivalence class of loops)
has some loop as a representative.
\paragraph*{Brouwer Fixed Point Theorem} The agents completed a proof of the Brouwer Fixed Point Theorem,
though ultimately it depends on an unproven theorem that
the fundamental group of the circle is isomorphic to the integers.
Assuming the fundamental group of the circle is isomorphic to the integers,
the agents have proven (with a 2390 line proof) that the inclusion map from the circle to
the real plane is not nulhomotopic (not homotopic to a constant map).
Using this result, the agents have proven (with a 3729 line proof)
that for every nonvanishing vector field on the closed unit disk,
there is a point on the circle that points directly inward
and a point on the circle that points directly outward.
From this the agents proved the Brouwer Fixed Point Theorem
using a 1564 line proof.

The agents were not yet able to prove the fundamental group of the circle
is isomorphic to the integers.
By manually inspecting the state of the formalization,
we suspect part of the difficulty comes from an unsatisfactory definition
of $\cos$ and $\sin$.
The two functions are defined using the epsilon choice operator
as the pair of continuous real functions $(\gamma,\sigma)$ such that
$\gamma(0) = 1$,
$\sigma(0) = 0$,
$\forall x\in \mathbb{R}.(\gamma(x))^2+(\sigma(x))^2 = 1$,
$\forall x y\in \mathbb{R}.\gamma(x+y) = \gamma(x)\gamma(y) - \sigma(x)\sigma(y)$ and
$\forall x y\in \mathbb{R}.\sigma(x+y) = \sigma(x)\gamma(y) + \gamma(x)\sigma(y)$.
The agents prove no properties of these functions and would need
to prove there exist such functions before even the defining properties
could be extracted. Unfortunately, the pair of functions is not uniquely
determined. The pair of functions $\cos(kx)$ and $\sin(kx)$ (for any integer $k$) would also satisfy the properties. In particular, the pair of the constant $1$ function and the constant $0$ function satisfy the properties.
Clearly the definitions of $\cos$ and $\sin$ are faulty
and trying to prove anything using them could only distract the agents.
The agents would need to generate new (correct) definitions and use those.

\bibliography{afgpt}

@inproceedings{hales2017formal,
  title={A formal proof of the Kepler conjecture},
  author={Hales, Thomas and Adams, Mark and Bauer, Gertrud and Dang, Tat Dat and Harrison, John and Hoang, Le Truong and Kaliszyk, Cezary and Magron, Victor and McLaughlin, Sean and Nguyen, Tat Thang and others},
  booktitle={Forum of mathematics, Pi},
  volume={5},
  pages={e2},
  year={2017},
  organization={Cambridge University Press}
}

@book{hales2012dense,
  title={Dense sphere packings: a blueprint for formal proofs},
  author={Hales, Thomas Callister},
  volume={400},
  year={2012},
  publisher={Cambridge University Press}
}

@misc{urban2026130klinesformaltopology,
      title={130k Lines of Formal Topology in Two Weeks: Simple and Cheap Autoformalization for Everyone?}, 
      author={Josef Urban},
      year={2026},
      eprint={2601.03298},
      archivePrefix={arXiv},
      primaryClass={cs.LO},
      url={https://arxiv.org/abs/2601.03298}, 
}

@inproceedings{DBLP:conf/nips/SchickDDRLHZCS23,
  author       = {Timo Schick and
                  Jane Dwivedi{-}Yu and
                  Roberto Dess{\`{\i}} and
                  Roberta Raileanu and
                  Maria Lomeli and
                  Eric Hambro and
                  Luke Zettlemoyer and
                  Nicola Cancedda and
                  Thomas Scialom},
  editor       = {Alice Oh and
                  Tristan Naumann and
                  Amir Globerson and
                  Kate Saenko and
                  Moritz Hardt and
                  Sergey Levine},
  title        = {Toolformer: Language Models Can Teach Themselves to Use Tools},
  booktitle    = {Advances in Neural Information Processing Systems 36: Annual Conference
                  on Neural Information Processing Systems 2023, NeurIPS 2023, New Orleans,
                  LA, USA, December 10 - 16, 2023},
  year         = {2023},
  url          = {http://papers.nips.cc/paper\_files/paper/2023/hash/d842425e4bf79ba039352da0f658a906-Abstract-Conference.html},
  bibsource    = {dblp computer science bibliography, https://dblp.org}
}

@inproceedings{DBLP:conf/iclr/YaoZYDSN023,
  author       = {Shunyu Yao and
                  Jeffrey Zhao and
                  Dian Yu and
                  Nan Du and
                  Izhak Shafran and
                  Karthik R. Narasimhan and
                  Yuan Cao},
  title        = {ReAct: Synergizing Reasoning and Acting in Language Models},
  booktitle    = {The Eleventh International Conference on Learning Representations,
                  {ICLR} 2023, Kigali, Rwanda, May 1-5, 2023},
  publisher    = {OpenReview.net},
  year         = {2023},
  url          = {https://openreview.net/forum?id=WE\_vluYUL-X},
  bibsource    = {dblp computer science bibliography, https://dblp.org}
}

@inproceedings{DBLP:conf/mkm/BrownP19,
  author       = {Chad E. Brown and
                  Karol Pak},
  editor       = {Cezary Kaliszyk and
                  Edwin C. Brady and
                  Andrea Kohlhase and
                  Claudio Sacerdoti Coen},
  title        = {A Tale of Two Set Theories},
  booktitle    = {Intelligent Computer Mathematics - 12th International Conference,
                  {CICM} 2019, Prague, Czech Republic, July 8-12, 2019, Proceedings},
  series       = {Lecture Notes in Computer Science},
  volume       = {11617},
  pages        = {44--60},
  publisher    = {Springer},
  year         = {2019},
  url          = {https://doi.org/10.1007/978-3-030-23250-4\_4},
  doi          = {10.1007/978-3-030-23250-4\_4},
  bibsource    = {dblp computer science bibliography, https://dblp.org}
}

@inproceedings{DBLP:conf/mkm/BrownKSU25,
  author       = {Chad E. Brown and
                  Cezary Kaliszyk and
                  Martin Suda and
                  Josef Urban},
  editor       = {Valeria de Paiva and
                  Peter Koepke},
  title        = {Hammering Higher Order Set Theory},
  booktitle    = {Intelligent Computer Mathematics - 18th International Conference,
                  {CICM} 2025, Brasilia, Brazil, October 6-10, 2025, Proceedings},
  series       = {Lecture Notes in Computer Science},
  volume       = {16136},
  pages        = {3--20},
  publisher    = {Springer},
  year         = {2025},
  url          = {https://doi.org/10.1007/978-3-032-07021-0\_1},
  doi          = {10.1007/978-3-032-07021-0\_1},
  bibsource    = {dblp computer science bibliography, https://dblp.org}
}

\end{document}